\documentclass[nofootinbib,11pt]{revtex4}
\usepackage{graphicx}
\usepackage{amsmath,amsfonts,amsthm,amssymb,epsfig}
\usepackage{multirow}
\newcommand{\beq}{\begin{equation}}
\newcommand{\eeq}{\end{equation}}

\newcommand{\beqa}{\begin{eqnarray}}
\newcommand{\eeqa}{\end{eqnarray}}

\newcommand{\bean}{\begin{eqnarray*}}
\newcommand{\eean}{\end{eqnarray*}}

\newcommand{\psiu}{\Psi}
\newcommand{\matter}{$\mu$atter}

\newcommand{\calH}{{\cal H}}

\begin{document}

\title{Space does not exist, so time can.}
\author{Fotini Markopoulou\\
Perimeter Institute for Theoretical Physics}
\date{December 1, 2008}

\begin{abstract}
It is often said that in general relativity time does not exist.  This is because the Einstein equations generate motion in time that is a symmetry of the theory, not true time evolution.  In quantum gravity, the timelessness of general relativity clashes with time in quantum theory and leads to the ``problem of time'' which, in its various forms, is the main obstacle to a successful quantum theory of gravity.  I argue that the problem of time is a paradox, stemming from an unstated faulty premise.  Our faulty assumption is that space is real.   I propose that what does not fundamentally exist is not time but space, geometry and gravity.  The quantum theory of gravity will be spaceless, not timeless.  If we are willing to throw out space, we can keep time and the trade is worth it.  

\end{abstract}

 \maketitle
 
 \section*{The {\em paradox} of time in quantum gravity}

There are two kinds of people in quantum gravity.  Those who think that timelessness is the most beautiful and deepest insight in general relativity, if not modern science, and those who simply cannot comprehend what timelessness can mean and see evidence for time in everything in nature.  What sets this split of opinions apart from any other disagreement in science is that almost no one ever changes their mind, there is practically no crossing camps on the issue of time.  On some days, this makes me wonder if the split is truly on scientific grounds or something deeper. 
But for the purposes of this essay, we will stay on the fairly hard science and I will argue that time exists and that 
the problem of time in quantum gravity should be seen as a paradox.  Paradoxes are usually resolved when we realize that a certain unstated assumption is lurking in the background which, under closer inspection, we see is false.  
\vskip 0.3cm

\noindent
{\em Timelessness.\/}
What is the reason to believe that time does not exist, despite our obvious experience to the contrary?  Many books have been written on this and I  will just state the bare facts.  

General relativity is the theory of spacetime geometry interacting with matter.  A spacetime is a 4-dimensional curved manifold with metric $g_{\mu\nu}$ and curvature $R_{\mu\nu}$.  Matter, for the purposes of relativity, can be succinctly squeezed into the stress-energy tensor $T_{\mu\nu}$. Both metric and matter  are dynamical and affect each other:  matter tells spacetime how to curve and spacetime tells matter where to go.  This is the content of the Einstein equations:
\begin{equation}
R_{\mu\nu}-\frac{1}{2}g_{\mu\nu}R=\frac{8\pi G}{c^4} T_{\mu\nu}.
\label{eq:EE}
\end{equation}
$G$ is Newton's constant and $c$ is the speed of light.   

There are two points to note here.  The first we already made, the metric $g_{\mu\nu}$ is a dynamical field in general relativity, this is a theory of spacetime itself.  The second is that $g_{\mu\nu}$ is not quite the correct physical quantity.  The Einstein equations are invariant under diffeomorphisms of the spacetime manifold, operations that map spacetime points to other spacetime points.  A spacetime point as labeled by a given $g_{\mu\nu}$ is not physical.  {\em Events} are physical, and events are marked by interactions of physical objects.
It is photons bouncing off the paper and hitting your retina while you read this essay that make this sentence be part of your past.  A given collection of events and the order in which they occurred is physically meaningful but it can be represented by several different metrics, all related by diffeomorphisms.   The correct physical quantity is the equivalence class of metrics under diffeomorphisms, usually called a {\em geometry}.

Now let us consider a pure gravity scenario, a universe with no matter, where the right hand side of equation (\ref{eq:EE}) is zero. 
With no physical objects to interact and mark events, 
 diffeomorphisms are allowed to shuffle spacetime points freely.  When I write the 3+1 version of equation (\ref{eq:EE}) -- where 
 a spatial geometry, the world at a single instant of time, evolves forward in time -- I find that time evolution is a diffeomorphism, that is, it does nothing.  The Hamiltonian for the evolution of space is just a constraint. 
I just discovered that time does not exist. 

Remember that we find that time does not exist in a universe of pure gravity.  If we allow matter, we can properly define clocks and time.  And we live in a universe with matter.  The problem is that general relativity allows for pure gravity solutions and there is no way to exclude them in the current setup.  We are stuck with timelessness. 

\vskip 0.3cm

\noindent
{\em The problem of time.\/}
Timelessness enters in center stage in quantum gravity when quantum gravity is stated as the problem of unifying general relativity and quantum theory.  Quantum theory always uses a fixed spacetime.  When I fire my laser in the lab there is a clock on the wall which is not involved in my experiment.  I do use my clock when I prepare my state and then measure it ten minutes later.  It enters the Schr\"{o}dinger equation as a parameter and allows me to predict the possible outcomes of my experiment.  But I don't look at the state of the laser to figure out what time it is, time is a prior notion, external to the quantum mechanical evolution.

Quantum gravity is often seen as the problem of unifying or reconciling general relativity and quantum theory, combining the physics of the very large with that of the very small.  There are certainly places where we need both, such as black hole physics, where no real progress can be made without a quantum theory of gravity.  The incompatibility of general relativity and quantum theory can be stated in many ways and a classic one is {\em the problem of time}:  adapting the Scr\"{o}dinger equation to a diffeomorphism invariant context by quantizing  equation (\ref{eq:EE}) gives the Wheeler-deWitt equation,
\begin{equation}
\widehat{H}|\psiu\rangle=0.
\label{eq:wdw}
\end{equation}
$|\psiu\rangle$ is the wavefunction of the universe, $\widehat{H}$ is the quantum Hamiltonian constraint and 0 means there is no time.  

The problem with this shortest of equations is that we still do not really know what it means.  After half a century of hard work, we have candidates for $|\psiu\rangle$ and $\widehat{H}$ but 0 is still causing trouble. The current version of the problem of time can be called  {\em  the low energy limit problem}.  Let us say you have a $|\psiu\rangle$ and $\widehat{H}$ that you think are the correct ones.  They belong to the fundamental quantum theory of gravity, that is, they decribe
 the world at extremely high energy.  How do you test they are correct?  There is only one way, to compare with known physics and experiment.  You must find the low energy limit of your $|\psiu\rangle$.  This is where 0 comes in.  Without a background geometry, 
how do we distinguish low energy from high energy?  Defining energy needs not just time but time translation invariance, so without time we don't even know what energy is, let alone what is low energy and high energy.

We now see what the problem of time controversy is: is timelessness an important lesson from general relativity that must be carried over to a quantum theory of gravity or
can we allow a microscopic theory with time?  
If it is the first, how do we get around the above problems?  And if it is the second, since 
any quantum theory of gravity must recover general relativity in the appropriate limit, how does the  time of the fundamental theory go away in the limit of pure gravity?

There are candidate quantum theories of gravity of both kinds.  Loop Quantum Gravity \cite{LQG} is a well-known example of an implementation of the Wheeler-deWitt equation and the main challenge it faces is indeed finding the low energy limit of the theory.  
A prominent example of a quantum theory of gravity with fundamental time is Causal Dynamical Triangulations \cite{CDT} which gets beautiful low energy results such as the correct 3+1 dimension of the universe as an emergent dynamical property.  The theory still has to show gravity in the appropriate limit, so we do not yet know how, and if, the fundamental time can be reconciled with the diffeomorphism invariance of general relativity.

Quantum gravity has been facing the problem of time for half a century.  Will it continue to cause trouble for the next half century?
Or is the problem of time a fake problem?  Could it be a paradox?  We arrived at it by putting together true statements from general relativity and from quantum theory and we found a contradiction.   Now, either fundamental science is stuck or we just have a paradox that will go away if we find the faulty premise.  What can the faulty premise be?

\section*{A world without geometry}

It is true that the view of quantum gravity as the unification of general relativity and quantum theory is in many ways becoming outdated.  First of all, for the first time observational data {\em is}, in fact, driving the field of quantum gravity.  Quantum gravity is not just about black hole physics anymore.  
Observational cosmology leaves us with 95\% of the universe unknown and lots of hard data on the high energy physics of the early universe.  Any quantum gravity theorist worth their salt must strive to make a connection between their theory and the data. 
What is dark energy and dark matter?  What is the inflaton and are there remnants of transplanckian physics in cosmology?  Is the scale of quantum gravity the Planck scale or something much larger? 
 Second, some kind of unification of general relativity and quantum theory is provided in a number of theories, from string theory to loop quantum gravity and causal dynamical triangulations.  The challenge for these theories is now to show they are the correct ones, again by making contact with the known world.  

It is then more appropriate to think of the problem of quantum gravity in broader terms:  one more situation in physics where we have the low energy theory -- general relativity and quantum field theory -- and are looking for the high energy, microscopic, fundamental one.  This appears pretty similar to things we have done before, for example, in going from thermodynamics to the kinetic theory.  Not surprisingly, this significant shift in perspective opens up new routes that may take us out of the old problems. 

\vskip .3cm

\noindent
{\em Geometrogenesis.\/}
We do not know much about quantum gravity.  We expect it to hold at Planck scale energies of $10^{22}$ MeV. 
This is $10^{20}$ orders of magnitude higher than subatomic physics.  Our best tests of quantum field theory 
are at settings of  some $10^{20}$ orders of magnitude cooler than the physics of quantum gravity.  In all other cases we know of in physics, it suffices to change energy scales by much less than that for the studied system to undergo a phase transition.  Then two things can happen.
First, the system can acquire properties that did not exist previously.  This is the subject of emergence.  
For example, if we cool water we get ice.  The order of the crystalline structure of ice is an emergent property of the ice phase. 
Second, but related, almost always the degrees of freedom we use to describe the two phases are different.  An example of this is the comparison of fluid dynamics to the underlying theory of quantum molecular dynamics.  The first is governed by the continuity and Euler equations and we use notions such as waves to describe the system.  At the molecular level we have the Schr\"{o}dinger equation and the intermolecular potential.  Waves are emergent notions, applicable only at the low energy phase.  There are no
degrees of freedom describing waves at the molecular level.  

It is reasonable to expect that quantum gravity will follow a similar pattern.  What we currently know is the low energy theory, the analogue of fluid dynamics.  We are looking for the fundamental theory, the analogue of the quantum molecular dynamics.   Just as there are no waves in the molecular theory,  we will likely not find geometric degrees of freedom in the fundamental theory.  By analogy with known physics, we should expect that the quantum theory of gravity is not a theory of geometry.   

I must emphasize that no geometry does not mean discrete or fuzzy geometry.  It means that the most primary aspects of geometry, such as the notion of ``here'' and ``there'' will cease to make sense.  
In fact, we have been grappling with no geometry for a while, in the traditional quantum gravity settings.  A quantization of general relativity leads to a quantum superposition of all geometries.  
What is not often appreciated is that the superposition of all geometries is nothing like a geometry.  
Take two geometries or, equivalently, two causal structures 1 and 2.   Let us say that in geometry 1 event A and event B are close while in geometry 2 they are far.  If you are event A, in world 1 you see event B  here, in world 2 you see event B  there.  Now superpose not only 1 and 2 but all possible geometries.  B is all over the place.  
 (This is the easy version, I am assuming you can identify A and B in both which, in fact, you generally cannot do since you need the notion of here and there to assign identity). 
  At Planck scale we are not in the semiclassical limit  and quantum geometry is not just the quantum fluctuation  of the lightcones of the classical world, it is 
 not geometry as we know it.
  The monstrosity we just created  does not even have a sensible notion of here and there, the most basic aspect of geometry.  It also does not have a notion of dimension \cite{DT}.  
 It's only the fact that we call it ``quantum geometry'', a combination of two words we understand, that fools us into thinking we comprehend it.

Let us name {\em geometrogenesis} the process of transitioning from a fundamental quantum theory of gravity without geometry to the known geometric physics.  
The geometrogenesis scenario is a simple one.  At high energy, or in the early universe,  there is no notion of geometry or geometric locality.   The system can be described in terms of microscopic, quantum degrees of freedom, which we call micro-matter, or \matter\ for short.  This is not our usual matter; by the same reasoning that we do not expect geometry to be valid more than twenty orders of magnitude below what we can probe, we also do not expect matter to be found intact at the quantum gravity scale.  At lower energies, or as the universe cools down, the system dynamically settles near its ground state.  This is the geometry plus matter phase, with ordinary matter and the symmetries that characterize the geometry of our deSitter universe.

\vskip .3cm

\noindent
{\em Information before geometry.\/}
Having raised the possibility that geometry does not exist at the fundamental level, we now need to find a way to do physics without geometry.  This may appear hard because all our physics is done with geometry.  But we can use a
relational and information theoretic language.  

As a first step, let us think of geometry not as an independent entity but as derivative from the properties of matter.  We turn around the usual order:  a particle is not Poincar\'{e} invariant because it is in a Minkowski spacetime, rather, all we can mean by a Minkowski spacetime is that all particles and their interactions are Poincar\'{e} invariant.  
Geometry is a concise way to describe the symmetries of the system. 
 The premise of geometrogenesis is that the
  symmetries that define geometry arise at low energy.  With this in mind, let
 us give a concrete model of geometrogenesis, specifically of emergent locality and
 emergent geometry, via symmetries of the ground state of a non-geometric quantum system. 

We consider a quantum and finite relational universe:  it contains $N$ constituents and only their relations are important.  This can be modeled as a network of $N$ nodes $a,b,...=1,...,N$ with a Hilbert space $\calH_{ab}$ attached to each link $ab$.  The states on $\calH_{ab}$ are the possible relations between $a$ and $b$.  The simplest case is that of two-state links  $\calH_{ab}=\{|0\rangle,|1\rangle\}$
representing locality or adjacency:  $|1\rangle$ state on link  $ab$ means the link is {\em on} and $a,b$  are {\em local}, while $|0\rangle$ means the link is {\em off}:
\[
\includegraphics{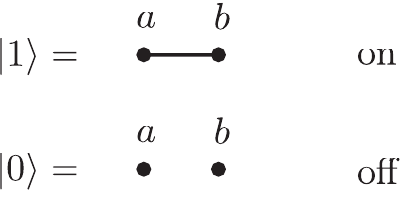}
\]
$K_N$ is the complete graph on $N$ vertices, i.e., the graph in which there is one edge connecting every pair of vertices, so that there is a total of $N(N-1)/2$ edges and each vertex has degree $N-1$. The total state space of our system is
$
\calH=\bigotimes^{\frac{N(N-1)}{2}}\calH_{ab}.
$
A generic state of $\calH$ is a superposition of subgraphs of $K_N$.
 A Hamiltonian operator $H$ will assign energy $E(G)$ to a graph $G$:
$
E(G)=\langle\Psi_G|H|\Psi_G\rangle.
$
The dynamics on this system performs a geometrogenesis transition from the high to the low energy phase.

In \cite{QGph,LR}, we studied a Hamiltonian that provides  an explicit model for geometrogenesis, with locality, translations, etc, being properties of the ground state.  In Table 1, we have summarized the properties of the model, which we call {\em Quantum Graphity}, at high and low temperatures.  At high energy, the dynamics is invariant under permutations of the vertices.  
There is no notion of locality, i.e., the entire universe is one-edge adjacent to any vertex.  Said differently, there is no notion of a subsystem, in the sense of a local neighborhood, since the neighborhood of any vertex is the entire $K_N$.  The microscopic degrees of freedom, \matter, are the states of the link spaces.  They evolve under the Hamiltonian  in a time parameter which plays the same role as lowering the temperature.

The ground state is a graph with far less edges than $K_N$.  For the purposes of this essay,  the naive geometric interpretation of the ground state graph as a lattice geometry, with distances given by graph geodesics, etc, will be sufficient to see that the ground state is a graph of very large diameter corresponding to a low-dimensional lattice.  Permutation invariance breaks to translations.
In addition, by enlarging the link state spaces and choosing an appropriate Hamiltonian we obtain matter, photons and fermions, in the ground state via the mechanism of string net condensation of Levin and Wen \cite{snets}.
Subsystems can be defined as subgraphs of the ground state or, better, as the emergent matter excitations, and the dynamics of the emergent matter is local.   

Quantum graphity is an explicit model of geometrogenesis.  In addition, in such a model, observable effects of emergent locality such as its imprint on the CMB can be studied \cite{LR}.   Finding the right dynamics for the desired ground state is ongoing work.  I do not want in this essay to discuss the merits and challenges of this model but just use the network based universe to illustrate geometrogenesis.

\begin{table}[htdp]
\caption{The two phases of quantum graphity.}
\begin{center}
\begin{tabular}{p{5cm}p{4cm}p{5cm}}
High energy&
\multirow{8}{*}{
$\begin{array}{c}
\mbox{geometrogenesis}\\
 \longrightarrow\\
\end{array}$
}
&Low energy\\
\\
\includegraphics[width=.2\textwidth]{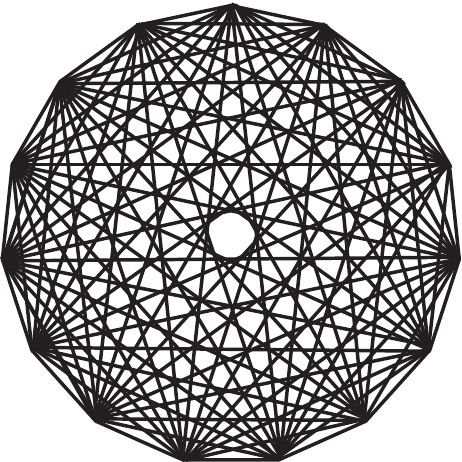}
&
&
\includegraphics[width=.2\textwidth]{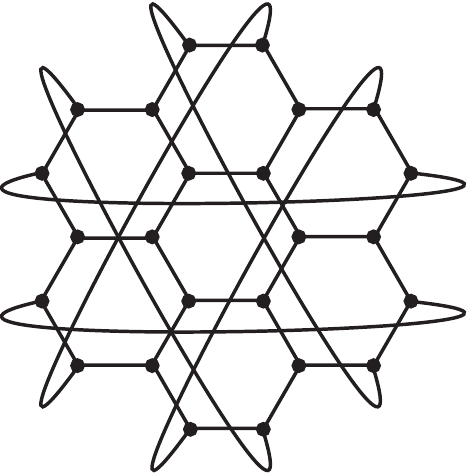}
\\
\\
$\bullet$ permutation symmetry&&
$\bullet$ translations\\
$\bullet$ no locality&&$\bullet$ local\\
$\bullet$ relational&&$\bullet$ relational\\
$\bullet$ no subsystems&&$\bullet$ subsystems\\
$\bullet$ external time&& $\bullet$ external and internal time\\
$\bullet$ $\mu$atter&&$\bullet$ matter + dynamical geometry
\end{tabular}
\end{center}
\label{default}
\end{table}%

In summary, we can use a dynamical network of quantum relations to describe a universe without geometry.  There is no need to assign any geometric properties to the network, it can simply be thought of as a quantum information processor.  In general, an information theoretic perspective is useful in studying emergence of geometry as it allows us to work with dynamical but ``disembodied'' (or ``degeometrized'') systems \cite{qi}.  

We have argued  geometry is not fundamental, instead quantum gravity is a theory of geometrogenesis.  We also illustrated geometrogenesis in  a concrete model.  With this in mind, what does our no geometry perspective tells us about time, diffeomorphisms and gravity?

\section*{Keeping track of time{\em s}}

It should now be clear that there are two possible notions of time:  
 the time related to the $g_{00}$ component of the metric describing the geometry at low energy and the time parameter in the fundamental microscopic Hamiltonian.  
Let us call the first {\em geometric time} and the second {\em fundamental time}.    
In our geometrogenesis context, it is clear that the geometric time will only appear at low energy, when geometry appears.   The problem of the emergence of geometric time is  the same as the problem of the emergence of space, of geometry.  
``Time does not exist'' refers to geometric time.  By making the geometry not fundamental, we are able to make a distinction between the geometric and the fundamental time, which opens up the possibility that, while the geometric time is a symmetry, the fundamental time is real.  
By distinguishing the two notions of time we may be able to have our cake and eat it: emerging geometric time from fundamental time is not remotely as intractable as dealing with a fully timeless world.  

It is important to note that the
relation between geometric and fundamental time is non-trivial and that the existence of a fundamental time does not necessarily imply a preferred geometric time.

\section*{We are inside the Universe}

The advantage, and the challenge, of the no-geometry perspective to quantum gravity is to understand how the Einstein equations -- gravity -- can arise in this new context.  This would complete the answer to the problem of time, and provide a quantum theory of gravity,  and 
is the final question I want to raise in this essay.  A  warning is due:  since we do not have the  completed quantum theory of gravity (yet), this part of the essay is meant to suggest and provoke a discussion rather than convince.  

\vskip .3cm

\noindent
{\em ``Faking'' independent geometric degrees of freedom.\/}
One can object to the statement that  fundamentally  geometry does not exist by pointing out that general relativity has independent geometric degrees of freedom.    
Doesn't quantum gravity have to have independent microscopic degrees of freedom corresponding to geometry?  

This is not true.  It is a basic principle of emergence that the collective is often not well described by the dynamics of the constituents.  It can appear to have a life of its own and can be described as if governed by rules that are independent of the microscopic constituents.  The lesson of ``more is different''  is that the collective may appear to have independent degrees of freedom. 

As an example, think what we mean by society.  We have capitalist societies, agricultural societies, totalitarian  societies.  We can describe the features of each and are not at all confused when we hear phrases such as ``our society is addicted to credit'' or ``society made him do X''.  But a society does not exist independent of its members. It is the actions, beliefs and expectations of people that form the society.   
To a certain extend, when things are in equilibrium we can study societies as entities of their own and ignore the people.  Out of equilibrium, in turbulent times,  we would certainly not be surprised when the actions of individuals are not negligible perturbations to the system but truly affect it.

While this analogy is not perfect, I am suggesting that we can see spacetime geometry as the analogue to society, with the role of individuals played by matter and its dynamics.  Near equilibrium, geometry can appear to have independent degrees of freedom.  However, that is not really the case, and we next argue that this gives rise to the Einstein equations.

\vskip .3cm

\noindent
{\em Gravity and diffeomorphisms.\/}
General relativity is a cosmological theory, meaning that it describes the entire universe.  In a cosmological theory all observables refer to information measurable by observers inside the universe.
In \cite{internal}, we demanded that this is the case and found that this requires modifying the algebra of observables so that they only give truth values in the backwards light cone.  In its quantum form, this means that the Wheeler-DeWitt equation is unphysical.  

We usually think of the Einstein equations as ``matter tells geometry where to go and geometry tells matter how to curve''.  In this essay we have taken a new step:  geometry is nothing but the collective organization of emergent matter.  
This leads to a new way to view the Einstein equations:  there is 
no surprise that  $T_{\mu\nu}$ and $R_{\mu\nu}$ are inter-related, they are different facets of the same thing.  In quantum graphity, \matter\ becomes  both geometry and matter.  This is a concrete realization of the  conjecture of \cite{OD}:  
Starting with a pre-geometric quantum system,  extract effective excitations and  use the same excitations to describe both the geometry metric field and curvature and the matter stress-energy tensor.  Such a unified approach will result in these two not being independent but satisfying the Einstein equations as identities (to lowest order).

If this conjecture is true, the task for the no-geometry perspective is to {\em explain} the Einstein equations and gravity rather than quantize. 
If this is indeed the true content of the Einstein equations,  then diffeomorphism invariance, the symmetry of the Einstein equations, is nothing more than the expression that we are trapped inside a system that appears to have a life of its own, but really its no more than our collective actions.

The disappearance of geometric time dictated by diffeomorphism invariance may be a statement about being inside the universe and not about whether fundamental time exists.   

\section*{We can keep time if we throw out space}

In this essay I argued that the problem of time is really a paradox that can be traced to taking spacetime geometry too seriously, beyond its domain of applicability.  
Timelessness refers only to the geometric time, not the microscopic fundamental time.  Fundamental time exists but space, geometry and gravity do not.  

In the discussion I clearly left out a number of issues that are essential to the nature of time, such as the role of observers, quantum mechanics or the arrow of time.  Partly this is because all of these are subtle and a sensible treatment would require more space than what is available here.  But the real reason is that, in my opinion, a non-geometric physics is essential to progress in these fundamental questions and 
all of these issues should be re-examined in the no-geometry context. 

The ultimate background independent theory is not one where the physics is independent of the background but the one with {\em no} background.  Background independence means to make no distinction between geometry and matter and to describe the dynamics of the universe as seen by observers inside it. 
Diffeomorphisms are not about timelessness but about being inside a dynamical universe, affecting it and being affected by it, constituting it.

\bibliographystyle{prsty}
\bibliography{bib}

\begin{thebibliography}{99}
\bibitem{LQG} 
Abhay Ashtekar, Jerzy Lewandowski, 
``Introduction to Modern Canonical Quantum General Relativity'', 
Class.Quant.Grav.21:R53,2004 , 
arXiv:gr-qc/0404018.
\bibitem{CDT} 
J. Ambjorn, J. Jurkiewicz, R. Loll,
``Quantum Gravity, or The Art of Building Spacetime'',
in "Approaches to Quantum Gravity - toward a new understanding of space, time, and matter", edited by D. Oriti, CUP 2009, 
arXiv:hep-th/0604212.
\bibitem{DT}
Jan Ambjorn, Bergfinnur Durhuus, Thordur Jonsson,
``Quantum Geometry : A Statistical Field Theory Approach'',
CUP 2005.
\bibitem{QGph} 
Tomasz Konopka, Fotini Markopoulou, Simone Severini,
``Quantum Graphity: a model of emergent locality'',
Phys.Rev.D77:104029,2008, arXiv:0801.0861.
\bibitem{LR}
Alioscia Hamma, Fotini Markopoulou, Isabeau Premont-Schwarz, Simone Severini,
``Lieb-Robinson bounds and the speed of light from topological order'',  Phys.Rev.Lett., in press,
arXiv:0808.2495.
\bibitem{snets}
Michael A. Levin, Xiao-Gang Wen, 
``String-net condensation: A physical mechanism for topological phases'', 
Phys.Rev. B71 (2005) 045110, arXiv:cond-mat/0404617.
\bibitem{qi}
Fotini Markopoulou, 
``New directions in Background Independent Quantum Gravity'', 
in "Approaches to Quantum Gravity - toward a new understanding of space, time, and matter", edited by D. Oriti, CUP 2009,
arXiv:gr-qc/0703097.
\bibitem{internal}
Fotini Markopoulou, 
``The internal description of a causal set: What the universe looks like from the inside'',
Commun.Math.Phys. 211 (2000) 559-583,
arXiv:gr-qc/9811053.
\bibitem{OD}
Olaf Dreyer,
``Emergent General Relativity'',
in "Approaches to Quantum Gravity - toward a new understanding of space, time, and matter", edited by D. Oriti, CUP 2009,
arXiv:gr-qc/0604075.

\end{thebibliography}

\end{document}